\documentclass[twocolumn,aps,floats,floatfix,prl,nofootinbib,superscriptaddress,tightenlines,preprintnumbers]{revtex4}
\usepackage{graphicx} \usepackage{bm}
\usepackage{amsmath}

\newcommand{\beq}{\begin{equation}}
\newcommand{\eeq}{\end{equation}}
\newcommand{\bea}{\begin{eqnarray}}
\newcommand{\eea}{\end{eqnarray}}
\newcommand{\dd}{\text{d}}

\begin{document}
\preprint{UCSD/PTH-06-06, UTTTG-06-06, CMUHEP-06-07}

\title{Falsifying Models of New Physics via  $WW$ Scattering}
\author{Jacques Distler}
\thanks{\uppercase{W}ork supported by NSF
contracts PHY-0071512 and PHY-0455649.}
\affiliation{University of Texas,
Dept.\ of Physics,
Austin,  Texas 78712, USA}
\author{Benjamin Grinstein}
\thanks{\uppercase{W}ork supported in
    part by \uppercase{DOE}
under contract DE-FG03-97ER40546.}
\affiliation{University of California, San Diego,
Dept.\ of Physics,
La Jolla, California  92093-0319, USA}
\author{Rafael A. Porto}
\thanks{\uppercase{W}ork supported by \uppercase{DOE}
contracts \uppercase{DOE-ER}-40682-143 and
\uppercase{DEAC02-6CH03000}.} \affiliation{Carnegie-Mellon
University, Dept.\ of Physics, Pittsburgh, Pennsylvania  15213,
USA}
\author{ Ira
Z. Rothstein}
\thanks{\uppercase{W}ork supported by \uppercase{DOE}
contracts \uppercase{DOE-ER}-40682-143 and
\uppercase{DEAC02-6CH03000}.}
\affiliation{Carnegie-Mellon University,
Dept.\ of Physics,
Pittsburgh, Pennsylvania  15213, USA}

\begin{abstract}
We show that the coefficients of operators in the electroweak chiral
Lagrangian can be bounded if the underlying theory obeys the usual
assumptions of Lorentz invariance, analyticity, unitarity and crossing
to arbitrarily short distances. Violations of these bounds can be
explained by either the existence of new physics below the naive
cut-off of the the effective theory, or by the breakdown of one of
these assumptions in the short distance theory.  As a corollary, if no light
resonances are found, then a measured violation of the bound would
falsify generic models of string theory.
\end{abstract}

\maketitle 

The standard model (SM) is only an effective field theory, a good
approximation only at energies below some scale $\Lambda$. This scale,
 however, is still undetermined. If, as
naturalness arguments indicate, new physics is required to explain the
relative smallness of the weak to Planck scale ratio, then we would
expect the theory to break down at energies of about 1~TeV. However, even if
naturalness arguments fail we still know that the SM, augmented by
Einstein gravity, must break down at the scale of quantum gravity,
where predictive power is lost.

In searching for low energy effects of the physics which underlies the
SM it is prudent to take the model independent approach of adding
operators of dimension higher than four to the SM Lagrangian and
parameterizing the new physics by their coefficients.  Dimensional
analysis dictates that these coefficients contain inverse powers of
$\Lambda$ so the precision with which we must extract them grows with
the scale of new physics. This decoupling phenomena makes falsifying
theories of the underlying short distance interactions (the
ultraviolet (UV)) extremely difficult. Indeed, if the scale of quantum
gravity is as high as the Planck scale, it becomes interesting to ask
the question as to whether or not the theory is, even in principle,
falsifiable. One possibility is that the mathematical structure leads
to unique low energy predictions.  However, in the case of  string
theory, recent progress seems to indicate that this is not a likely
scenario. Another possibility is that there are low energy, non-Planck
suppressed, consequences of some underlying symmetries. Symmetries
link the UV and the infrared (IR) by distinguishing between
universality classes.  However, string theory does not seem to have
any problems generating the low energy symmetries manifested at
energies presently explored.  Indeed, given the enormous number of
string vacua it may be that string theory can accommodate whatever new
physics is found at the TeV scale by the Large Hadron Collider (LHC).

Thus it seems that decoupling  may have the effect of rendering
string theory unfalsifiable. However, dispersion relations can be
used to establish bounds which, if violated, imply that the
underlying theory can not obey the usual assumptions  of Lorentz
invariance, crossing, unitarity and  analyticity. This type of
bound was raised a long time ago in the context of chiral
perturbation theory  \cite{Pham,Ananthanarayan:1994hf,portoles} and was
recently revisited in \cite{adams}. In this letter we will show
that such assumptions in general lead to bounds on the values of
coefficients of higher dimension operators in the SM \footnote{This possibility was
raised in \cite{adams}.}. As we shall see, the utility of these
bounds depends upon the value of the Higgs mass.

\begin{figure*}[t]
\begin{center}
\includegraphics[width=3.5in]{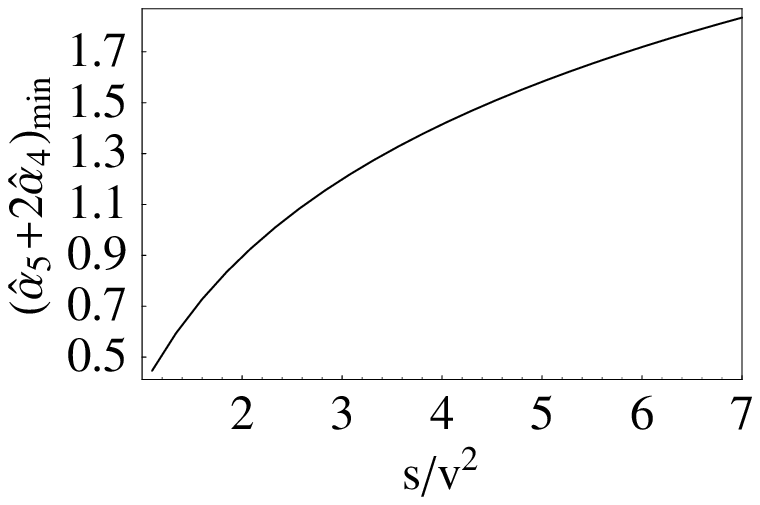}\hfill
\includegraphics[width=3.5in]{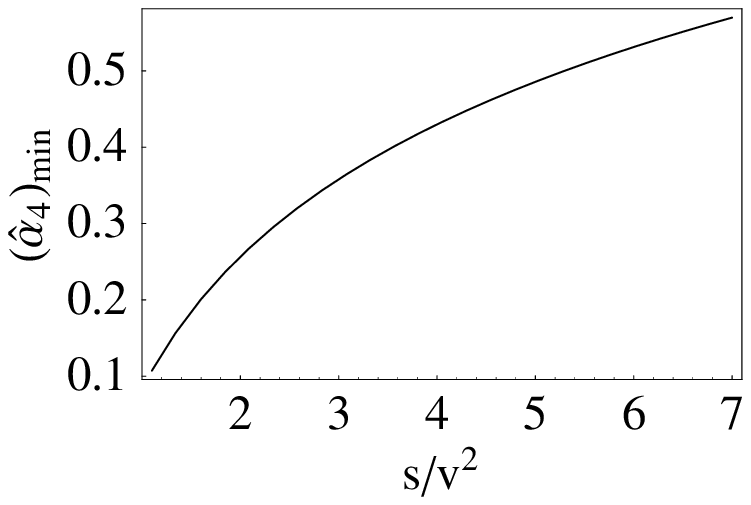}
\end{center}\vskip-20pt
\caption{\label{fig:alpha45min} Bounds on electroweak chiral parameters from $Z^0_LZ^0_L$
 and $ W_L^+Z_L^0$ scattering as a function of $s$ in the dispersion relation,
  Eq.~\eqref{naivebound}. }
\end{figure*}

In the absence of a light Higgs particle, symmetry considerations
dictate that the electroweak symmetry breaking sector of the SM be described by
a chiral Lagrangian of the nonlinearly realized spontaneously broken
$SU(2)\times U(1)$.  We will derive bounds on certain parameters in the
Lagrangian which are not well constrained from oblique corrections.
For simplicity we will assume that the strongly coupled dynamics
responsible for electroweak symmetry breaking preserves a custodial
$SU(2)$ symmetry.  This assumption, which is empirically validated by
the fact that the $\rho$ parameter is so close to unity, drastically
reduces the number of terms in the effective Lagrangian. The
Lagrangian we consider \cite{Appelquist:1993ka} contains, in addition
to the usual field strength terms for the electroweak gauge bosons, a
derivative expansion in the $SU(2)$ nonlinear sigma model fields,
\begin{multline}
\label{eq:EWlag}
{\cal L}={\cal
L}_{\text{gauge}}-{\textstyle\frac14}v^2\text{Tr}(V_\mu V^\mu)
+{\textstyle\frac{1}{2}}\alpha_1 gg^\prime \text{Tr}\left(B_{\mu \nu}T W^{\mu\nu}\right)\\
+{\textstyle\frac{1}{2}}i\alpha_2 g^\prime \text{Tr} \left( T
[V^\mu , V^\nu]\right)B_{\mu \nu}
+i \alpha_3 g \text{Tr} \left( W_{\mu \nu} [V^\mu,V^\nu]\right)\\
+\alpha_4 \left(\text{Tr}(V_\mu V_\nu) \right)^2 +\alpha_5\left(
\text{Tr}(V_\mu V^\mu)\right)^2
\end{multline}
where $T\equiv 2 \Sigma T^3 \Sigma^\dagger$, $V_\mu  \equiv (D_\mu
\Sigma)\Sigma^\dagger$ and $D_\mu\Sigma
=\partial_\mu\Sigma+{\textstyle\frac12}igW^a_\mu\tau^a\Sigma
-{\textstyle\frac12}ig'B_\mu \Sigma \tau^3$.
with $\Sigma(x)=\exp(i\pi^a(x)\tau^a/v)$ and $\tau^a$ the Pauli
matrices. The ``pion'' fields here play the role of the would-be
Goldstone bosons arising from the broken gauge symmetry. Had we
not imposed the custodial symmetry we would have included six
additional operators.  The coefficient $\alpha_1$ is strongly
constrained by virtue of its contribution to the gauge boson self
energies at tree level \cite{LEP}. The coefficients $\alpha_2$ and
$\alpha_3$ contribute at tree level to the anomalous three gauge
boson vertices which have been studied at LEP \cite{LEP}. Given
the constraints on these parameters, they will not be considered
in our analysis, as their effect on our bounds are small, although
their inclusion is straightforward. The final two coefficients,
$\alpha_4$ and $\alpha_5$, contribute to two to two scattering at
tree level and thus  bounds on them arising from loop corrections
to the $T$ parameter are rather weak \cite{Dawson:1998yi}. It is these
coefficients that we bound below.


The bounds on these couplings are obtained by considering longitudinal
$ZZ \to ZZ$ and $WZ\to WZ$ scattering. Assuming Lorentz invariance,
analyticity and unitarity the forward scattering amplitude $T$
satisfies the twice subtracted dispersion relation
\begin{multline} 
\label{naivebound} \frac{\dd^2 \hat T(s)}{\dd s^2} = 2! \int_{4m^2}^\infty 
\frac{\dd x}{\pi } \sqrt{x(x-4m^2)} \\ \times \left( \frac{\sigma(x)}{(x-s)^{3}} +
\frac{\sigma_u(x)}{(x-4m^2+s)^{3}}\right).
\end{multline} 
Here and below, $s$, $t$ and $u$ are Mandelstam variables.  We have
used the optical theorem to express the discontinuity across the cut
in terms of the scattering cross section $\sigma$ and the crossed channel
cross section $\sigma_u$. For simplicity we have denoted by $m$ the mass of
the scattering particles, but more precisely $2m^2$ should be replaced
by $2m_Z^2$ and $m_Z^2+m_W^2$ for $ZZ\to ZZ$ and $ZW\to ZW$,
respectively. We have introduced $T=\hat T +$~pole term, where the
pole term arises from the exchange of a gauge boson. The s-channel
poles are of the form $p(s)/(s-m^2)$, where $p$ is a polynomial. Since
the degree of $p$ is at most 3, the pole cancels from both sides of a
twice subtracted dispersion relation. The t channel poles vanish upon
differentiation. Note that the existence of a long range force renders
the charged particle scattering total cross section divergent (a pole
at $t=0$). Our bounds will rely only on interactions which contain no
Coulomb singularities. To obtain bounds we will use the Equivalence
Theorem (ET) to approximate the scattering amplitude of longitudinally
polarized massive vector bosons, $\hat T$, by that of pseudo-Goldstone
bosons~\cite{Cornwall:1974km,Lee:1977eg,ET}. The ET has been studied
extensively. It is by now well understood, in a loop expansion, how an
amplitude for scattering of (longitudinal) vector particles in a gauge
theory can be reproduced by that of pseudo-scalars in a non-linear
sigma model~\cite{Bagger:1989fc,Veltman:1989ud,ET2,ET3} to leading
order in an expansion in $m^2/s$ and $g^2$.

The ET approximation is valid provided $s\gg m^2$. Hence we take $s\sim
v^2+i0$ in the dispersion relation.  In this regard we deviate from
the classical analysis, which takes $s$ below threshold, $s<4m^2$, and
real. We then break the integral in Eq.~\eqref{naivebound} into two
terms, the integrals from $4m^2$ to $kv^2$ and from $kv^2$ to $\infty$. For
the latter we use the fact that the cross section is positive
definite, while the former is computed using the ET to evaluate the
cross section.

The constant $k$ is chosen to minimize the error introduced by our
approximations while keeping the right side of \eqref{naivebound}
positive. One loop electroweak corrections to the amplitude scale as
\beq
\label{radcor}
\delta_{ew}T\propto O\left(\frac{g^2s}{(4\pi  v)^2}\ln(s/\mu^2) \right)
\eeq
while chiral corrections scale as
\beq
\delta_\chi T \propto O\left(\frac{s^3}{(4\pi)^4 v^6}\ln(s/\mu^2) \right).
\eeq
Hence the optimal choice of $k$ should be a number of order unity. It
is easy to see that this can be achieved while keeping the right side
of \eqref{naivebound} positive. In fact, calculating $\pi^0\pi^0$ and $\pi^0\pi^+$
scattering ($Z^0_LZ^0_L$ and $Z^0_LW^{+}_L$ scattering in the ET approximation) we find
that the values of $k$ for which the real part of the integrals from
$4m^2$ to $kv^2$ vanish,
\begin{multline}
\text{Re}\left[ \int_{4m^2}^{kv^2} \frac{\dd x}{\pi } \sqrt{x(x-4m^2)} \right.\\
\left. \times \left( \frac{\sigma(x)}{(x-s)^{3}} +
\frac{\sigma_u(x)}{(x-4m^2+s)^{3}}\right)\right]=0,
\end{multline}
at a fixed value of $s+i0$, are well fit by
\bea
k&=&5.1(s/v^2)-0.4 \\
k&=&5.0(s/v^2)-0.2
\eea
for $\pi^0 \pi^0$and $ \pi^0\pi^+$ scattering respectively. Restricting $k$ so
that $kv^2\log(s/ \mu^2)/(4\pi v)^2\lesssim \delta $ determines how large $s$ may be
taken in the dispersion relation while keeping the errors from the
chiral expansion under control. For our numerical estimates we take
$\delta=1/5$ and explore the range from $\delta=1/2\times (1/5)$ to $2\times(1/5)$.

Bounds on $\alpha_4,\alpha_5$ follow from positivity of the right hand side of
the dispersion relation.  The left hand side of (\ref{naivebound}) may
be approximated using the ET. We use the results for one loop pion
scattering calculated in Ref. \cite{Gasser:1983yg}. Up to second order in the chiral
expansion,
\beq 
\label{eq:Tgiven} 
\frac{\dd^2\hat T}{\dd
s^2}(s)=\frac1{24\pi^2v^4}\left[\sum_{i=1,2} c^i \hat \alpha_i +
\tilde T(s)\right], 
\eeq 
The expressions for the $c^i,\tilde T$ depend on which physical
amplitude is considered.  The coefficients $\hat \alpha_{4(5)}$ are defined in
a similar fashion as the $\bar \ell_i$ in \cite{Gasser:1983yg}, except that the
renormalization point is taken at $\mu=v$:
\begin{equation}
\label{a}
\alpha_i^r(\mu)=\frac{\gamma_i}{96\pi^2}\left[\hat \alpha_i+\frac14\ln(v^2/ \mu^2)\right]\,,
\end{equation}
with $\gamma_5=1$ and $\gamma_4=2$.

\begin{figure}[t]
\begin{center}
\includegraphics[width=3.5in]{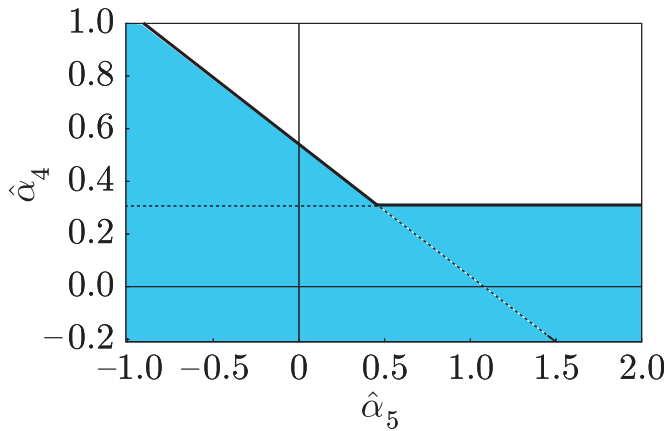}
\end{center}
\caption{\label{fig:pipibound} Bounds on the coefficients of the
  electroweak chiral lagrangian, $\hat \alpha_4$ and
$\hat \alpha_5$ renormalized at the scale $v$.}
\end{figure}

It is now straightforward to obtain the bounds by computing the
real part on the LHS of (\ref{naivebound}).  Fig.~\ref{fig:alpha45min}
shows the bounds from $Z^0_LZ^0_L$ and $W^{+}_LZ^0_L$  scattering for the chosen value of
$s$. The best bound (largest $s$) is obtained by allowing $k$ as large
as allowed by the restriction on chiral corrections,
$k\log(s/\mu^2)/16\pi^2\lesssim\delta $,  at $\mu=v$. We
find 
\begin{align}
\label{bound45}\hat \alpha_5+2 \hat \alpha_4 \; &\geq 1.08
\\
\label{bound4}
\hat \alpha_4 \; &\geq 0.31
\end{align}
These are the bounds which correspond to $\delta =1/5$. If we vary $\delta$ by a
factor of 2 or 1/2 the first bound changes by +0.37 and $-0.35$, while
the second by +0.13 and $-0.12$. Note that we consider the choice $\delta
=1/5$ to be quite conservative since the integral from $kv^2$ to $\infty$,
which we have neglected, is positive definite. The bounds are shown in
the $\hat \alpha_4$ vs $\hat \alpha_5$ plane in Fig.~\ref{fig:pipibound}. It
should be kept in mind that while the fractional uncertainty in our
bounds seem large, the relevant parameters for $WW$ scattering is the
renormalized coupling constant at a scale comparable to the $W$ and
$Z$ masses. The large uncertainty quoted above corresponds to 25\% and 21\%
uncertainty in the bounds for the renormalized couplings at
$\mu^2=4m_W^2$.  The dominant error on these bounds is due to
electroweak loop corrections whose contributions are down by
$g^2v^2/s\approx 0.4g^2$, where the last equality is from the
numerical value of $v^2/s$ obtained by restricting the chiral
corrections to be smaller than $\delta=1/5$.  These corrections, which
are not included in \eqref{bound45}--\eqref{bound4}, are unlike the
chiral corrections in that they are calculable and will appear in a
subsequent publication\cite{dgpr}. Alternatively we can estimate the
uncertainty introduced by our approximations by re-computing the
bounds retaining the subleading terms in $m^2/s$ in the pion
scattering amplitude. The result is to move down the bound on $\hat
\alpha_5+2 \hat \alpha_4 $ in \eqref{bound45} to 1.04 while the bound on $\hat \alpha_4
$ in \eqref{bound4} stays at 0.31, consistent with the error estimates
above.

In the future it may be possible to measure these coefficients through
measurements of $WW$ or $ZZ$ scattering at the LHC or the NLC. Studies
suggest a sensitivity to $(\alpha_4,\alpha_5)$ in $WW$ and $ZZ$ scattering at a
linear collider that could well establish a result in contradiction
with our bounds \cite{Kilian:1996fk,Boos:1997gw}. The bounds can also
be tested in QCD. In the limit $g,g'\to0$ the EW chiral lagrangian of
\eqref{eq:EWlag} reduces to the hadronic chiral
lagrangian\cite{Gasser:1983yg}, and the parameters of the former,
$\alpha_{4,5}$, tend towards those of the latter, $\ell_{1,2}$. Working above
threshold, $s>4m^2_\pi$, as we have done here, is probably not
trustworthy for bounds on parameters of the hadronic chiral lagrangian
since in that case $m_\pi/ f_\pi \gtrsim 1 $ (as compared with $m_W/v\lesssim 1/3$) and
thus the validity of the chiral approximation used for the right hand
side of the dispersion relation comes into question. Nevertheless,
bounds on $\bar l_1$ and $\bar l_2$ derived from
\eqref{bound45}--\eqref{bound4} using $\bar l_1=4\bar \alpha_5$ and $\bar
l_2=4 \bar\alpha_4$, with $\bar \alpha_i=\hat\alpha_i+\frac14\ln(v^2/m^2)$ are
consistent with the experimental values quoted in \cite{gb}. It is not
necessary to choose $s$ above threshold in the dispersion relation to
bound $\ell_{1,2}$, since there is no need to invoke the ET to perform
the calculation.  The optimum bound obtained by working below
threshold and for non-forward scattering, $t>0$\cite{Martin},
reproduces the somewhat weaker bounds found in \cite{portoles}.

Ref.~\cite{adams} has proposed that constraints on $\ell_{1,2}$ can also
be obtained by requiring the absence of superluminal
propagation. When the chiral Lagrangian is expanded
about the classical background  $\Sigma=\exp(ic\cdot x \tau^3)$,
for some constant vector, $c^\mu$, the absence of superluminal
excitations gives
$\ell^{\text{cl}}_2 > 0$, 
$\ell^{\text{cl}}_1 + \ell^{\text{cl}}_2 > 0$. 
Classical propagation in a nontrivial background is tantamount to
studying forward scattering off that background and, for this process,
chiral loops are generally as important as the tree-level contributions
of $\ell_{1,2}$.  Including chiral loop corrections shifts the bounds from
forward scattering: up  $\bar\ell_2\geq (39\pi-92)/48$ and down
 $\bar\ell_1+2\bar\ell_2\geq(9\pi-36)/32$.  Moreover, a third bound 
appears,  $\bar\ell_1+ 3 \bar\ell_2\geq0.91$, 
changing the \emph{shape} of the excluded region.
Note that these are not the strongest bounds obtainable on $\bar\ell_{1,2}$.
Stronger bounds can be obtained from dispersion relations in the unphysical
($t\to 4m^2$) regime (see, {\it e.g.}, \cite{portoles}). But they do
demonstrate the unreliability of the classical approximation. Neglecting
the chiral loops is tantamount to making an
additional assumption about the underlying UV theory,
namely, that it is weakly-coupled. This is an unwarranted
assumption about the nature of the UV physics since it
cannot be justified from considerations of the low-energy
effective theory alone.

Let us now consider the implications of our new bounds.  Suppose
that no light Higgs is found and the bounds are violated.  There
are then two logical possibilities.  Either the cut-off is lower
than expected, $\Lambda\ll 4\pi v$, and the subsequent power
corrections, of order $(s/ \Lambda)^2$ invalidate the bounds, or
the underlying theory has an $S$ matrix which does not have the
usual analytic properties we associate with causal, unitary
theories.  The former possibility is what we would expect if the
underlying strong dynamics leading to electroweak symmetry
breaking were a large-$N$ gauge theory. In the large-$N$ limit the
masses of resonances are suppressed by $1/N$ (holding the
confinement scale fixed) and, as such, the cut-off is effectively
reduced. The masses of the resonances would have to be
sufficiently light to invalidate the bounds.  It is interesting to
note that this is exactly the situation one would expect in a
Randall-Sundrum scenario where the gravitational theory is dual to
a large-$N$ gauge theory.  In principle one could retool the
bounds in this case,  by including the resonances in the effective
theory thus raising the cut-off scale. One could then test whether
this new effective theory is the low energy limit of a theory with
an analytic $S$-matrix. In the absence of a light Higgs or other
light resonances, a violation of the bound on
$(\hat \alpha_4,\hat \alpha_5)$ would indicate a breakdown of one or more
basic properties of the $S$-matrix. The assumptions used in
obtaining the dispersion relation are: Lorentz invariance: the
amplitude can only depend upon the three Mandelstam invariants.
Analyticity and crossing: the cuts lie on the real axis as shown
in the figure, with no singularities on the physical sheet off the
real axis. Unitarity:  the imaginary part of the scattering
amplitude along the cuts is positive. String theory, which is
designed to be valid at all distance scales,  is constructed to
produce an $S$-matrix with precisely these properties. More
generally, if the bounds are violated, whatever underlying
dynamics is responsible for the electroweak chiral Lagrangian must
not satisfy these basic properties of $S$-matrix theory. 
Theories which could violate the bound include those 
which violate Lorentz invariance \cite{Mattingly}, or unitarity \cite{Gambini}.

 There remains,
however, the question of what energy scale the new physics (which
violates one or more of the above assumptions) enters and what the
nature of that new physics is. It is tempting to assert that the
scale of the unconventional new physics should  not be too far
above the cutoff of the electroweak chiral Lagrangian,
$\Lambda\sim4\pi v$. But, absent a better characterization of the
nature of this new physics (which, by definition, differs from
that of conventional quantum field theory or string theory), it
would be hard to present a proof of that assertion.

\begin{acknowledgments}
We would like to thank Nima Arkani-Hamed and Aneesh Manohar for discussions.
\end{acknowledgments}

\end{document}